%% file: mass_metal_igm.tex
\documentclass[useAMS,usenatbib]{mn2e}
\usepackage{color}
\usepackage{multirow}

\include{pacchetti}


\include{definizioni}

\def\msun{{\rm M}_{\odot}}
\def\zsun{{\rm Z}_{\odot}}
\def\vir{\rm vir}
\def\halo{\rm h}
\def\cent{\rm c}
\def\citeSim{P14}
\def\rvir{r_{\vir}}
\def\zlos{\langle Z\rangle}
\def\HI{\hbox{H~$\scriptstyle\rm I $~}}

\newcommand\textlcsc[1]{\textsc{\MakeLowercase{#1}}}

\begin{document}
\date{}
\pagerange{\pageref{firstpage}--\pageref{lastpage}} \pubyear{2014}
\title[The CGM of high-$z$ galaxies]{The circumgalactic medium of high redshift galaxies}
\author[Pallottini et al.]{A. Pallottini$^{1}$, S. Gallerani$^{1}$, A. Ferrara$^{1,2}$\\
$^{1}$Scuola Normale Superiore, Piazza dei Cavalieri 7, I-56126 Pisa, Italy\\
$^{2}$Kavli IPMU, The University of Tokyo, 5-1-5 Kashiwanoha, Kashiwa 277-8583, Japan}

\maketitle
\label{firstpage}

\begin{abstract}
We study the properties of the circumgalactic medium (CGM) of high-$z$ galaxies in the metal enrichment simulations presented in \citet[][]{Pallottini:2014_sim}. At $z=4$, we find that the simulated CGM gas density profiles are self-similar, once scaled with the virial radius of the parent dark matter halo. We also find a simple analytical expression relating the neutral hydrogen equivalent width (${\rm EW}_{\rm HI}$) of CGM absorbers as a function of the line of sight impact parameter ($b$). We test our predictions against mock spectra extracted from the simulations, and show that the model reproduces the ${\rm EW}_{\rm HI}(b)$ profile extracted from the synthetic spectra analysis. When compared with available data, our CGM model nicely predicts the observed ${\rm EW}_{\rm HI}(b)$ in $z\lsim2$ galaxies, and supports the idea that the CGM profile does not evolve with redshift.
\end{abstract}

\section{Introduction}
The circumgalactic medium (CGM) is the extended interface between the interstellar medium (ISM) of a galaxy and the surrounding intergalactic medium (IGM). This component plays a key role in galactic evolution as it represents a mass reservoir and a repository of the mechanical and radiative energy produced by stars. Due to its low density, the CGM can be almost uniquely traced by absorption line experiments towards background sources, typically quasars. The intervening CGM associated with a foreground galaxy then leaves a characteristic spectral feature. Provided that a sufficiently large sample of galaxies are available it is then possible to statistically determine the Equivalent Width (${\rm EW}$) of a given absorption line as a function of the line of sight (l.o.s.) impact parameter ($b$).

The CGM has been probed so far up to $z\sim3$ using absorption lines of both \HI \citep[e.g.][]{Rudie:2012ApJ,Rudie:2013ApJ,Pieri:2013arXiv} and heavy elements \citep[e.g.][]{Steidel:2010ApJ,Churchill:2013arXiv,Nielsen:2013ApJ,Borthakur:2013ApJ,Liang:2014arXiv}. These observations show that the CGM extends up to $b\simeq10\,\rvir$, where $\rvir$ is the virial radius of the parent dark matter (DM) halo. An anticorrelation between ${\rm EW}$ and $b$ is observed; moreover, the ${\rm EW}$ profiles appear to be self-similar once scaled with $\rvir$. Finally \citet{Chen:2012MNRAS} suggested that CGM absorption profiles show no signs of evolution from $z\simeq2$ to $z\simeq0$.

In the framework of a $\Lambda$CDM\footnote{Hereafter we assume a $\Lambda$CDM cosmology with $\Omega_{\Lambda}= 0.727$, $\Omega_{dm}= 0.228$, $\Omega_{b}= 0.045$, $\rm H_0=100~h~km~s^{-1}~Mpc^{-1}$, $\rm h=0.704$, $n=0.967$, $\sigma_{8}=0.811$ \citep[][]{Larson:2011}.} cosmological model, the CGM properties can be derived from numerical simulations simultaneously accounting for both large scale ($\simeq$~Mpc) structure and small scale ($\simeq$~kpc) galactic feedback. While such a huge dynamical range makes a truly self-consistent simulation impossible, these difficulties can be overcome by following the unresolved physical scales with subgrid models.

Along these lines, some numerical studies have focused on testing CGM metal enrichment models \citep[e.g.][]{Shen:2013ApJ, Barai:2013MNRAS, Crain:2013MNRAS}; others have investigated the imprint of the last phases of reionization on the IGM/CGM \citep[e.g.][]{Finlator:2013MNRAS, Keating:2013arXiv} or the ISM/CGM overdensity-metallicity ($\Delta$-$Z$) relation as a function of redshift \citep[][hereafter \citeSim]{Pallottini:2014_sim}. Surprisingly, little attention has been devoted so far to understand the physics beneath the observed CGM profile self-similarity and redshift independence.

In this Letter we show that the previously found $\Delta$-$Z$ relation naturally arises from self-similar nature of the CGM density/metallicity profiles. We use this result to derive an analytical expression for ${\rm EW_{\rm HI}}(b)$ which we then test against synthetic spectra extracted from the simulations and available observational data.
\section{Numerical simulations}
We adopt the simulations described in \citeSim~ which were obtained by using a customized version of the Adaptive Mesh Refinement code \textlcsc{ramses} \citep{Teyssier:2002}. Starting from cosmological initial conditions generated at $z=199$, we evolve a $(10$~Mpc~$h^{-1})^{3}$ volume until $z=4$. The DM mass resolution is $6.67 \times 10^{5}\,h^{-1}\msun$, and the adaptive baryon spatial resolution ranges from $19.53\,h^{-1}$~kpc to $1.22\,h^{-1}$~kpc.

We include subgrid prescriptions for star formation, accounting for supernova (thermal) feedback and implementing metal-dependent stellar yields and return fractions. Our simulation reproduces the observed cosmic star formation rate \citep[][]{Bouwens:2012ApJ} and stellar mass densities \citep[][]{Gonzalez:2011} for $4\leq z \lsim 10$.

We define a galactic environment as a connected patch of enriched gas with metallicity exceeding a chosen threshold, i.e. $Z > Z_{\rm th}\equiv10^{-7}\zsun$. Within such regions, following a common classification, we identify three different phases according to their gas overdensity: IGM ($\Delta = \rho/\bar{\rho} \leq 10$), CGM ($10<\Delta\leq 10^{2.5}$), and ISM ($\Delta> 10^{2.5}$).
%

In \citeSim~we have shown how to construct a complete catalogue of galactic environments at a given redshift. To each galactic environment we associate the group of DM halos, of total mass $M_{\halo}$, inside its boundary. We denote as ``central halo'' the most massive halo in each group and ``satellites'' the remaining ones. Since the central halo dominates the local dynamics, we use its mass ($M_{\cent}$) to compute the virial radius of the structure ($\rvir$).

\subsection{Self-similar $\Delta$ and $Z$ profiles}\label{sec_selfsim}
\begin{figure}
\centering
\includegraphics[width=8.7cm]{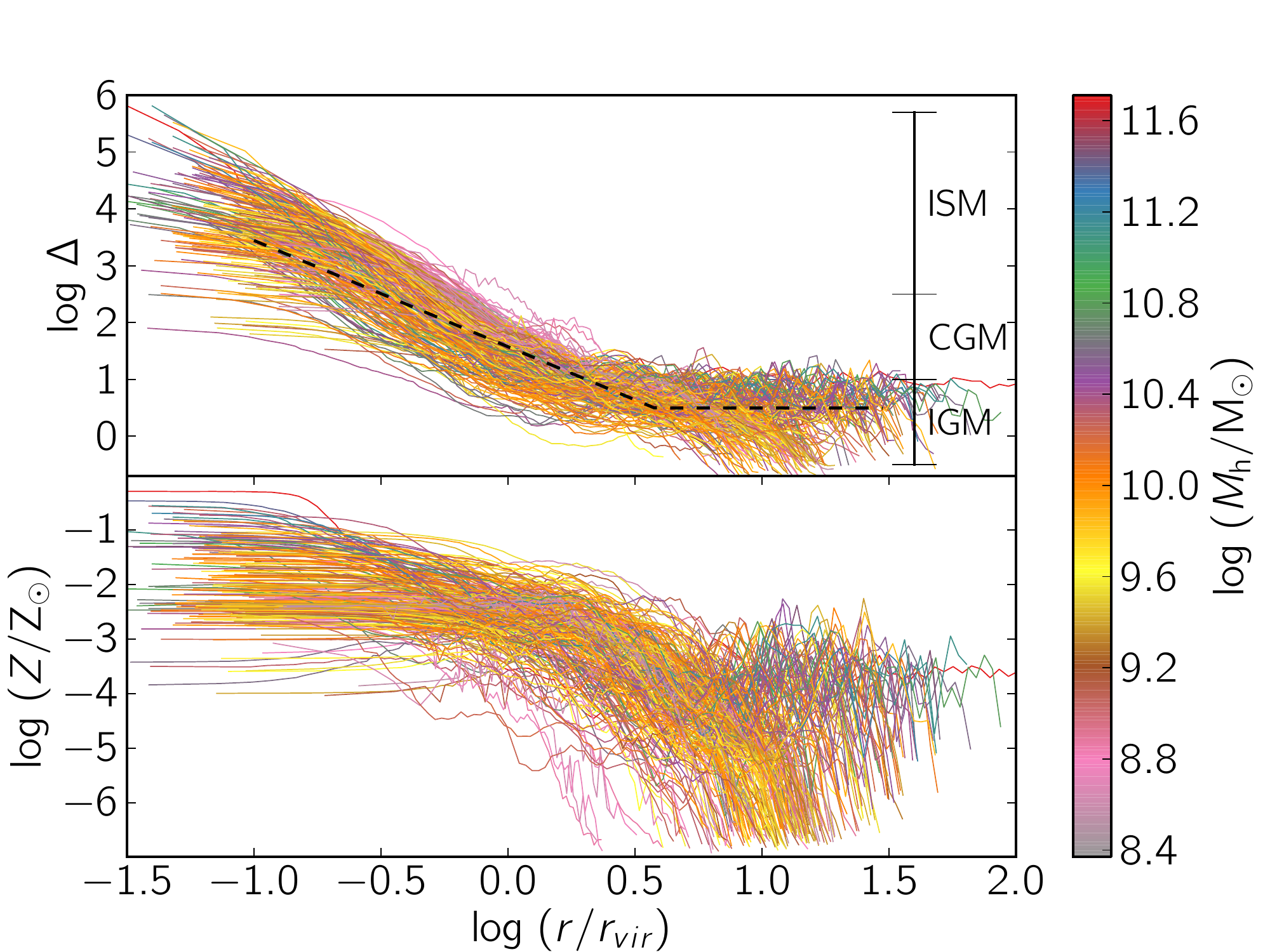}
\caption{Overdensity ($\Delta$, upper panel) and metallicity ($Z$, lower panel) radial profiles of $\simeq300$ simulated galactic environments at $z=4$. Each line refers to a selected environment, and its color corresponds to the associated total DM halo mass ($M_{\halo}$), as shown in the colorbar. The distance is normalized to $\rvir$, the virial radius of the central halo. In the upper panel the black dashed line is the analytical fit to the data (see eq. \ref{eq_densita}).
\label{delta_z_radial_profile}}
\end{figure}
We focus our attention at $z=4$, the lowest redshift reached by the simulation. At this epoch a $\Delta$-$Z$ relation for the gas is already in place. Fig. \ref{delta_z_radial_profile} shows the spherically-averaged radial profiles of the overdensity (upper panel) and metallicity (lower panel), as a function of $x\equiv r/\rvir$, for various ($\simeq300$) simulated galactic environments characterized by $10^{8.5}\lsim M_{\halo}\lsim10^{11.5}\msun$.

The overdensity trend is very similar for all galactic environments. In particular, the ISM is located at $x\lsim 10^{-0.5}$, the CGM spans the radial range $10^{-0.5}\lsim x\lsim 10^{0.5}$ and the IGM extends beyond $x \simeq 10^{0.5}$. Hence, the adopted density definition for different gas phases is equivalent to a distance classification\footnote{Therefore we will use interchangeably overdensity and distance definitions for the three phases.}, similarly to the proposal by \citet[][]{Shull:2014arXiv}.

Generally, one might expect that the shape of the relation depends somewhat on feedback prescriptions \citep[][]{Oppenheimer:2012MNRAS}. However, we note that feedback has been accurately calibrated in \citeSim~to reproduce globally averaged galactic properties.

The gas density profile can then be written in terms of the self-similar variable $x$ as a piecewise power-law of index $\alpha$:
\begin{equation}\label{eq_densita}
\rho_{\rm pp}(x)\slash\rho_{\rm vir}=\Theta(x_{\rm IGM}-x) x^{-\alpha}+\Theta(x-x_{\rm IGM})x_{\rm IGM}^{-\alpha}\,,
\end{equation}
where $\rho_{\rm vir}$ is the gas density evaluated at the virial radius, $\Theta$ is the Heaviside function and $x_{\rm IGM}$ denotes the location where the density approaches a constant value typical of the IGM in the proximity of galactic systems. The best fit values for the parameters are: $\alpha=1.87\pm0.05$, $\rho_{\rm vir}/\bar{\rho}=37.5\pm 4.7$ and $x_{\rm IGM} = 3.8\pm 0.2$. From the fit, we also find the relation $(\rho_{\rm vir}/\bar{\rho}) x_{\rm IGM}^{-\alpha}\simeq 3$. In the upper panel of Fig. \ref{delta_z_radial_profile}, $\Delta_{\rm pp}=\rho_{\rm pp}/\bar{\rho}$ is shown by the black dashed line.

The metallicity (Fig. \ref{delta_z_radial_profile}, lower panel) is essentially flat in the ISM and rapidly declines in the CGM; this result is independent from the selected environment. The trend is however not universal in the IGM, as expected from the results of \citeSim, where we showed that $Z$ is only weakly correlated with $\Delta$ in the IGM. Environments associated with massive DM halos ($M_{\halo}\gsim10^{9.5}\msun$) show an enriched ($Z\simeq10^{-3.5}\zsun$) IGM up to $x \simeq10^{1.5}$. Instead, halos with $M_{\halo}\lsim10^{9.5}\msun$, which contain small galaxy groups or even isolated galaxies, only manage to pollute the IGM within $x\lsim 10$. This trend is discussed in detail in Fig. 13 of P14.
\begin{figure}
\centering
\includegraphics[width=8.7cm]{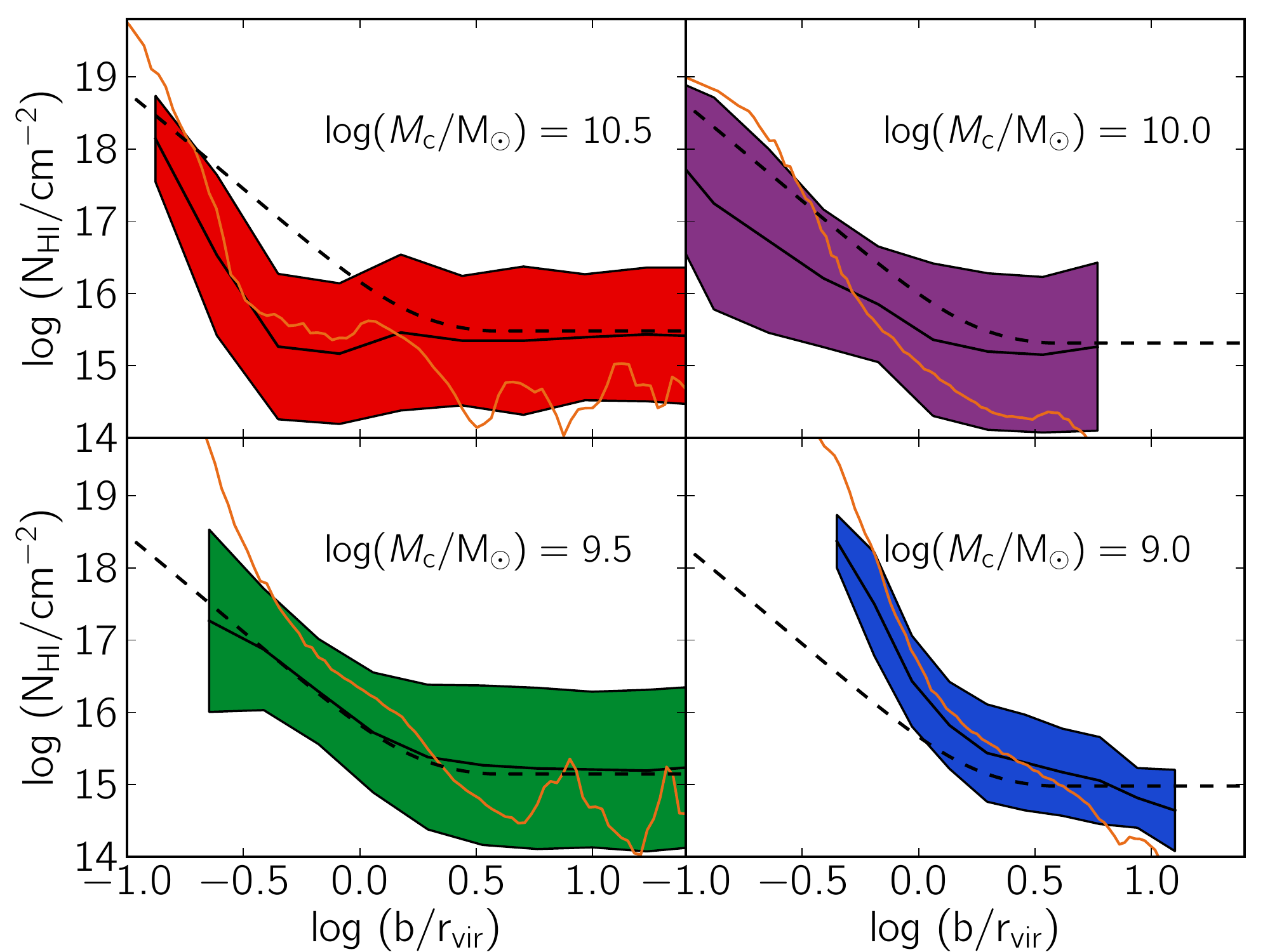}
\caption{Neutral hydrogen column density ($N_{\rm HI}$) as a function of the normalized impact parameter ($b/\rvir$) for a set of galactic environments with central halo mass $M_{\cent}$. The mean and the r.m.s. $N_{\rm HI}$ values inferred from the absorption spectra are shown through solid black lines and colored shaded regions, respectively. The solid orange lines represent the average $N_{\rm HI}$ inferred from the simulation (see footnote \ref{footnote4}), while the dashed black lines denotes the $N_{\rm HI}$ resulting from the analytical model (eq. \ref{eq_nh}) calculated for $\log(T_{\rm eff}/{\rm K})=5$.
\label{z4_nh_from_spectra}}
\end{figure}

Finally, galaxies hosted in halos with masses lower than $\simeq10^{8.5}\msun$ only show up as satellites (\citeSim). Their effect is perceivable in Fig. \ref{delta_z_radial_profile} as a local perturbation to the global $\Delta$ and $Z$ trends at $x \gsim 1$. The satellite positions resulting from our simulation are in broad agreement with the outcome of the numerical simulation by \citet{Khandai:2014arXiv}, who find a flat satellite distribution at $x \gsim 1$ for $M_{\halo}\sim10^{10}\msun$ (see their Fig. 10).

\subsection{Modeling \HI absorption}\label{sec_nh_plus_am}
From the fit to the simulated density profile, $\rho_{\rm pp}(x)$, we build a simple analytical model that describes the \HI absorption properties (N$_{\rm HI}$ and EW$_{\rm HI}$) of the CGM/IGM.

The \HI column density along a l.o.s. is defined as $N_{\rm HI}=\int n\,x_{\rm HI} \, {\rm d}l$, where $n$ is the total hydrogen density, and $x_{\rm HI}$ is the \HI fraction. Assuming spherical symmetry, we express $N_{\rm HI}$ through the following relation:
\be\label{eq_nh}
N_{\rm HI}(b)=\frac{2}{m_{\rm H}}\int_{b}^{l_{\rm max}} \rho_{\rm pp}x_{\rm HI} \frac{r}{\sqrt{r^{2}-b^{2}}}\, {\rm d}r\, ,
\ee
where $\rho_{\rm pp}$ is given in eq. \ref{eq_densita}, $b$ is the impact parameter, $m_{\rm H}$ is hydrogen mass, $l_{\rm max}=\sqrt{b^{2}+(\Delta v/H)^{2}}$, $H=H(z)$ is the Hubble constant and $\Delta v$ is the velocity window sampled by observations. Assuming local photoionization equilibrium \citep[e.g.][]{Dayal:2008MNRAS}, the \HI fraction can be written as $x_{\rm HI}=(1+\xi)-\sqrt{(1+\xi)^{2}-1}$, where $\xi\equiv(\Gamma_{\rm HI}m_{\rm H})/(2\,\rho_{\rm pp} \alpha_{\rm rec})$, $\alpha_{\rm rec}$ is the recombination rate and $\Gamma_{\rm HI}=\Gamma_{\rm HI}(z)$ is the UV background photoionization rate. Consistent with \citeSim~simulations, we use the UV intensity from \citet[][]{Haardt:2012}.

The \HI Ly$\alpha$ equivalent width can be expressed as follows:
\be\label{eq_EW}
{\rm EW}_{\rm HI}(b)=\frac{c}{\nu_{0}^{2}}\int_{0}^{\infty} \left(1-\exp(-N_{\rm HI} \sigma_{0} \phi)\right)\, {\rm d}\nu\, ,
\ee
where $c$ is the speed of light, $\nu_{0}$ is the frequency of the Ly$\alpha$ transition, $\phi=\phi((\nu-\nu_{0})/\Delta\nu,\Delta\nu)$ is the Voigt profile \citep[e.g.][]{Meiksin:2009RvMP}, $\Delta\nu=(\nu_{0}/c)\sqrt{2K_{B}T/m_{\rm H}}$ is the thermal Doppler broadening and $\sigma_{0}=\pi e^{2} f /m_{e} c$, where $f$ is the oscillator strength, $e$ and $m_{e}$ are the electron mass and charge, respectively. By combining eq.s \ref{eq_densita}-\ref{eq_EW}, we obtain the trend of ${\rm EW}_{\rm HI}$ with $b$, for different values of $\rvir$ and $T$.

The ${\rm EW}_{\rm HI}(b)$ dependence on $\rvir$, entering through the density dependence on $x=r/\rvir$, results in a stretching/compression of the density profile. The temperature $T$, entering in the expressions for $\alpha_{\rm rec}$ and $\Delta\nu$, regulates both \HI at $x\gg x_{\rm IGM}$ and the slope of the ${\rm EW}_{\rm HI}$ profile for $x\lsim x_{\rm IGM}$. For increasing (decreasing) $T$ values, \HI is shifted downward (upward) while the slope becomes steeper (shallower). It is worth noticing that we are assuming a single temperature value both for the IGM and CGM. Moreover, we are neglecting turbulence, which may affect the Doppler broadening of CGM absorbers \citep[e.g.][]{Iapichino:2013MNRAS}. Therefore, $T$ must be regarded as an ``effective temperature''; to make it clear we will use $T_{\rm eff}$ to indicate this quantity.
\section{Testing the \HI absorption model}\label{sec_spettri}
\begin{figure*}
\centering
\includegraphics[width=8.7cm]{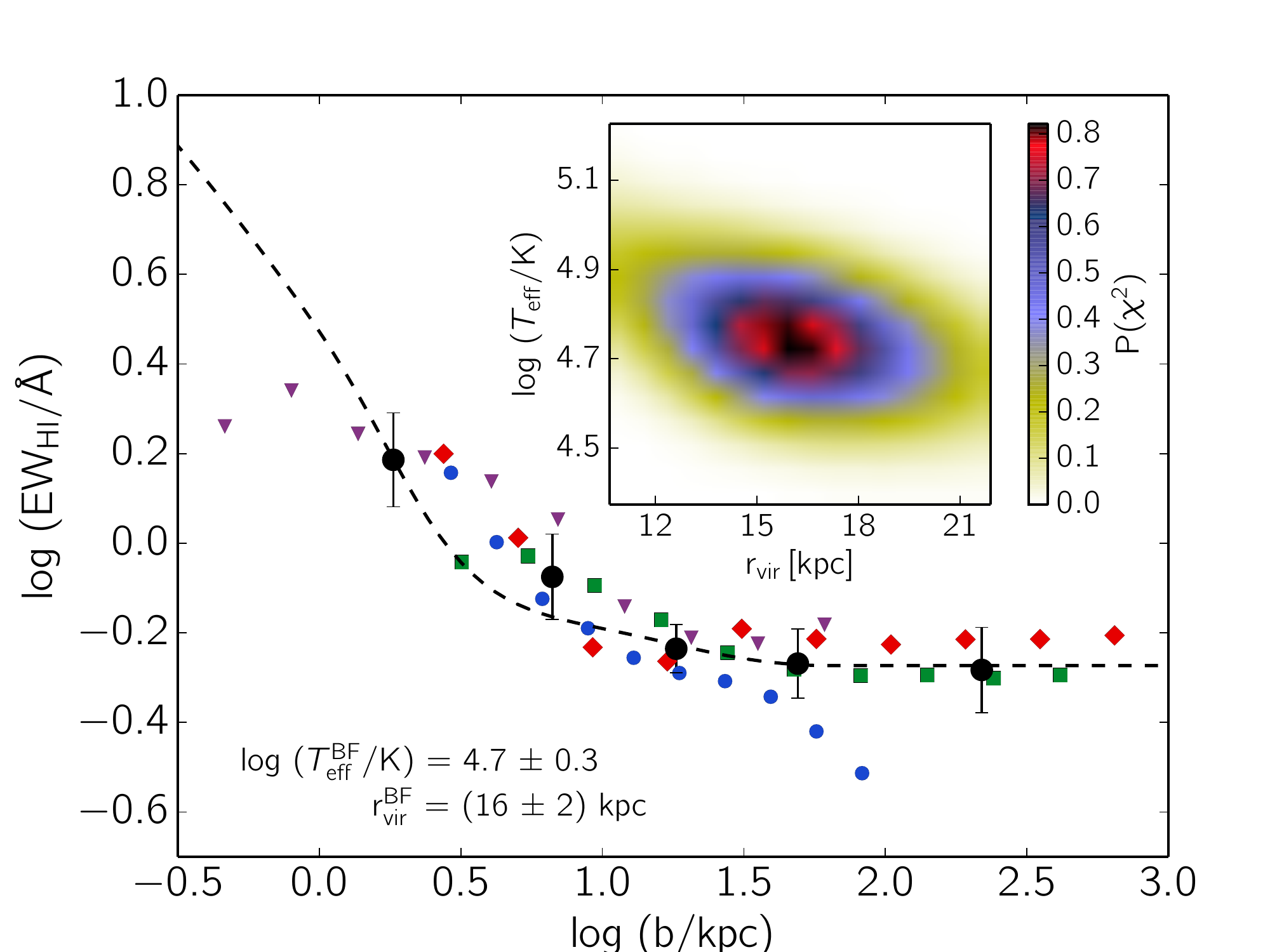}
\includegraphics[width=8.7cm]{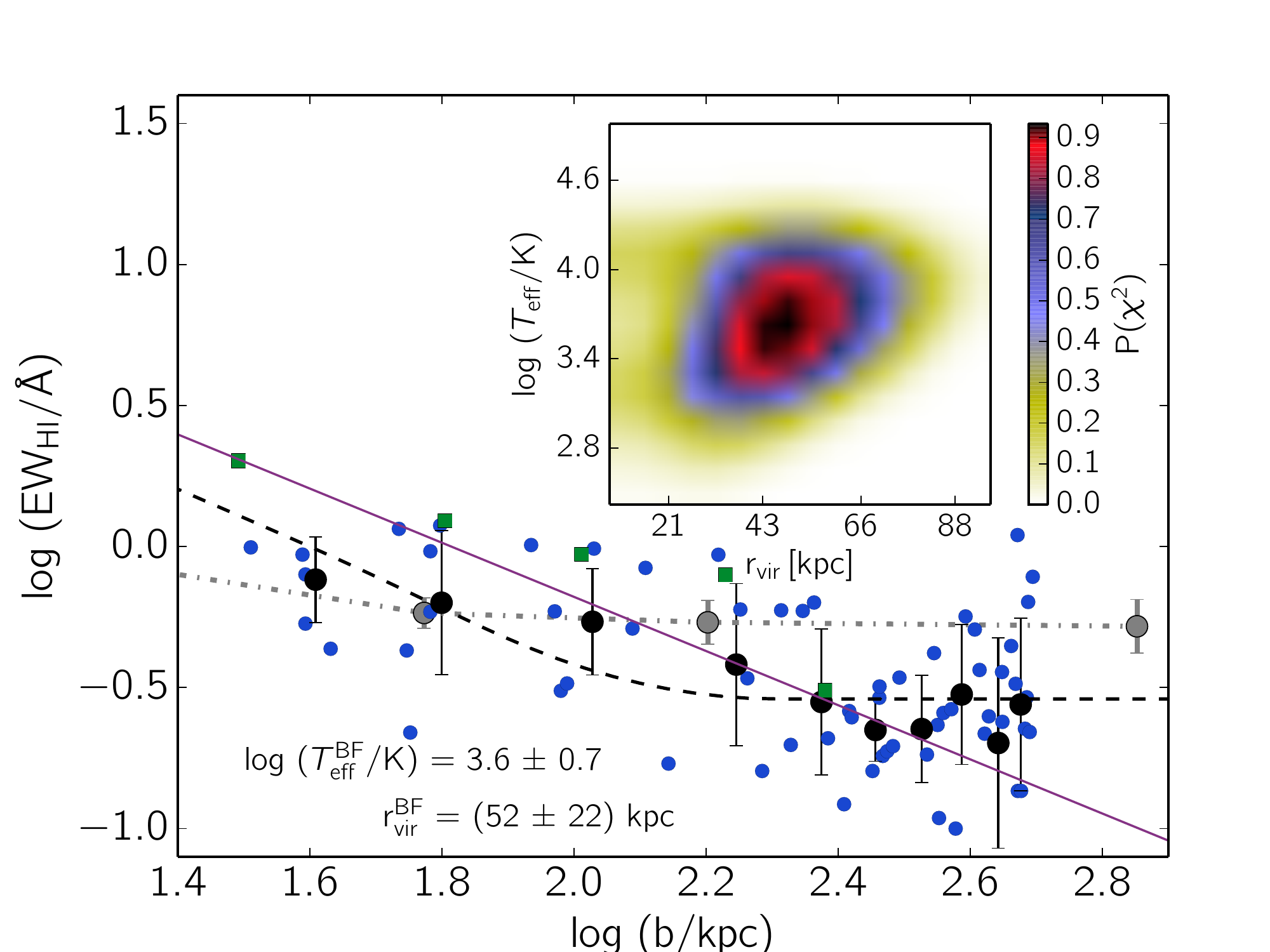}
\caption{Neutral hydrogen equivalent width (${\rm EW}_{\rm HI}$) profiles. {\bf Left panel}: \textit{simulated vs analytical profiles}. Simulated profiles for different galactic environments are indicated through red diamonds, purple downwards triangles, green squares, and blue circles for $M_c/M_{\odot}=10^{10.5}, 10^{10}, 10^{9.5}$ and $10^9$, respectively. Black circles with errorbars indicate the rebinned simulated data (see the text for the details). The black dashed line indicates the best fit analytical model, whose parameter ($T_{\rm eff}^{\rm BF}$ and $\rvir^{\rm BF}$) values are given. The inset shows the $\chi^{2}$ probability of the analytical model (eq.s \ref{eq_densita}-\ref{eq_EW}) as a function of $T_{\rm eff}$ and $\rvir$. {\bf Right panel}: \textit{simulated vs observed profiles}. Observations from \citet[][$z\simeq0.01$, blue circles]{Liang:2014arXiv} and \citet[][$z\simeq2.2$, green squares]{Steidel:2010ApJ}. The solid violet line represents the model by \citet{Chen:1998ApJ,Chen:2001ApJ}, and the grey circles linked by the dashed-dotted line show the rescaled ${\rm EW}_{\rm HI}(b)$ profile obtained from synthetic spectra at $z = 4$ (see the text for details). Additional notation is as in the left panel.
\label{ew_from_spectra_double}}
\end{figure*}
We are now ready to test our model both against simulated QSO absorption spectra and real data. 
\subsection{Synthetic \HI absorption spectra}
We compute mock QSO absorption spectra along several l.o.s. drawn through the simulated box. The technique adopted to compute the \HI optical depth is detailed in \citet{Gallerani:2006MNRAS}. In order to reproduce the mean transmitted flux observed at $z=4$ ($F_{\rm mean}=0.41$, Becker et al. 2013) the intensity of the UV ionizing background \citep[][]{Haardt:2012} assumed in the simulation is rescaled upwards by a factor 6.6, resulting into a photoionization rate $\log(\Gamma_{\rm HI}/{\rm s})=-12$. We also include observational artifacts in our simulated spectra, following \citet[][]{Rudie:2013ApJ}, a work based on HIRES spectra. We smooth the synthetic spectra to a resolution $R=45000$, add to each pixel a Gaussian random deviate, yielding a signal-to-noise ratio $S/N=100$, and we finally rebin the simulated transmitted flux in channels of width 0.4~A.

Among the l.o.s. extracted from the simulations, we select the ones passing through a specific galactic environment, defined by its central halo $M_{c}$ and its corresponding $\rvir$. The sample of l.o.s. considered encompasses a wide range of impact parameters, namely $10^{-1}$-$10^{2}$ $\rvir$.

\subsection{Largest gap statistics}

Along each l.o.s, we identify the CGM absorption feature with the largest spectral gap\footnote{Spectral gaps are defined as contiguous regions of the spectrum having an optical depth $\tau > 2.5$ over rest-frame intervals $>1$ A \citep{Croft:1998tsra.conf}.} found in the corresponding synthetic spectrum \citep[][]{Gallerani:2008MNRASa,Gallerani:2008MNRASb}. In order to check that largest gaps correctly identify CGM absorption features, we compute the $N_{\rm HI}$ along their corresponding l.o.s. paths, for different $b$ values, and for a set of galactic environments characterized by $M_{\cent}/\msun=10^{10.5}\,,10^{10}\,,10^{9.5}$ and $10^{9}$, that correspond at $z=4$ to $\rvir/{\rm kpc}=6.5, 9.6, 14$ and $21$, respectively.

The results from a sample of 3000 l.o.s. are shown in Fig. \ref{z4_nh_from_spectra}. Superimposed to the $N_{\rm HI}$ profiles obtained from the synthetic spectra analysis, are the average $N_{\rm HI}$ values inferred from the \citeSim~simulations\footnote{\label{footnote4} For each l.o.s., $N_{\rm HI}=\int n\, x_{\rm HI} \, {\rm d}l$, where the integration limits correspond to the borders of the environment, which in turn depend on $Z_{\rm th}$. Changing the metallicity threshold marginally affects the results: using $Z_{\rm th}=10^{-6}\zsun$ yields a variation of $\delta N_{\rm HI}\lsim 10^{14} {\rm cm}^{-2}$ for the inferred column density. The orange lines are obtained by averaging $N_{\rm HI}$ over $\sim10^{5}$ l.o.s..} and the column density resulting from the analytical model (eq. \ref{eq_nh}) calculated for $\log(T_{\rm eff}/{\rm K})=5$. Fig. \ref{z4_nh_from_spectra} shows that the largest gap statistics properly identifies CGM absorption features in \HI absorption spectra.

This technique is particularly promising for studying the CGM absorption properties for high-$z$ ($z>4$) quasar spectra. At these redshifts, the maximum observed transmitted flux drops below 50\% in the Ly$\alpha$ forest \citep{Songaila:2004AJ}. The resulting large uncertainties in the continuum determination may therefore hamper a proper Voigt profile analysis.

\subsection{Comparison with simulations}
As a next step, we compute the ${\rm EW}_{\rm HI}$ of the absorption features identified through the largest gap statistics as a function of the $b$ parameters, for the galactic environments presented in Fig. \ref{z4_nh_from_spectra}. The results are shown in the left panel of Fig. \ref{ew_from_spectra_double} through red diamonds, purple downwards triangles, green squares, and blue circles for $M_c/M_{\odot}=10^{10.5}, 10^{10}, 10^{9.5}$ and $10^9$, respectively. In the same figure, black circles and corresponding error bars represent mean and r.m.s. values obtained by averaging the ${\rm EW}_{\rm HI}$ profiles of the 4 different galactic environments into bins of width $\delta b$ such that $\log(\delta b/{\rm kpc}) \simeq 0.5$. As expected (see eq. \ref{eq_EW}), the ${\rm EW}_{\rm HI}$ profiles follow the $N_{\rm HI}$ trend (Fig. \ref{z4_nh_from_spectra}), namely ${\rm EW}_{\rm HI}$ decreases with $b$.

Finally, we fit the averaged ${\rm EW}_{\rm HI}$ profile resulting from the synthetic spectra analysis through our analytical model, finding the following best fit parameters: $\rm log (T_{\rm eff}^{\rm BF}/K)=4.7\pm 0.3$ and $\rvir^{\rm BF}=16\pm 2$~kpc. The inferred $T_{\rm eff}^{\rm BF}$ value agrees with typical values of the CGM/IGM temperature (\citeSim) and $\rvir^{\rm BF}$ is consistent with the average virial radius of the galactic environments considered, namely $\rvir^{\rm mean}=13\pm 6$~kpc. This result represents a solid consistency check of our model which allows us to repeat the same experiment on real data.

\subsection{Comparison with observations}

In the right panel of Fig. \ref{ew_from_spectra_double}, we compare our model with observations. Blue circles are the ${\rm EW}_{\rm HI}$ derived at $z=0.01$ by \citet[][]{Liang:2014arXiv}; black circles and corresponding error bars represent mean and r.m.s. values obtained by averaging the same observational data into bins $\sim 40$~kpc large.

For this comparison the model is calculated with $\log(\Gamma_{\rm HI}/{\rm s})=-13$, i.e. the value at $z=0.01$ given by \citet[][]{Haardt:2012}. By fitting\footnote{Supported by the lack of evolution of CGM profiles from $z=2$ to $z=0$ \citep[][]{Chen:2012MNRAS}, we assume $\rho_{\rm pp}$ to be redshift independent in eq. \ref{eq_densita}.} observations with our analytical model we find $\log(T_{\rm eff}^{\rm BF}/K)=3.6\pm 0.7$ and $\rvir^{\rm BF}=52\pm 22$~kpc. $T_{\rm eff}^{\rm BF}$ provides only an indicative value for the average temperature of the CMG/IGM. Although $\rvir^{\rm BF}$ is consistent within 1.2$\,\sigma$ with the mean virial radius $r_{\rm mean}=144\pm74$~kpc quoted by \citet{Liang:2014arXiv}, we note that our model favors smaller $\rvir$ values. 

The dashed black line in the right panel of Fig. \ref{ew_from_spectra_double} represents our best fit model, while the violet solid line shows the model from \citet[][hereafter C-model]{Chen:1998ApJ,Chen:2001ApJ}. Both the C-model and our best fit are in agreement with \citet{Liang:2014arXiv} observations up to $b/{\rm kpc}\sim 10^{2}$, i.e. in the CGM range ($b\sim4\,\rvir^{\rm BF}$). On the other hand, for $b/{\rm kpc}\gsim 10^{2.5}$, the C-model declines more steeply than data, while our best fit model properly reproduces the observed flat trend.

The gray circles linked by the dashed dotted line show the EW$_{\rm HI}(b)$ profile obtained from $z=4$ synthetic spectra, once rescaled to the $\rvir^{\rm BF}$ of the model which reproduces $z=0.01$ observations. The good agreement between the synthetic and observed EW$_{\rm HI}$(b) profiles, both shows that our modelling of the CGM reproduces observations, and favors the scenario suggested by \citet[][]{Chen:2012MNRAS} of a redshift independent CGM profile. As a further support to this idea we also show (green squares) CGM observations at $z\simeq2.2$ by \citet[][]{Steidel:2010ApJ}, which are perfectly consistent both with $z=0.01$ observations and with the EW$_{\rm HI}(b)$ profile obtained from $z=4$ synthetic spectra.
\section{Projected $\Delta$-$Z$ relation}
Inspired by the $\Delta$-$Z$ relation found in the ISM/CGM, we investigate whether the mean metallicity along a simulated l.o.s. $\zlos$ correlates with the $N_{\rm HI}$ distribution of our galactic environments. We compute $\zlos$ through the following equation: $\zlos =N_{\rm HI}^{-1} \int n\,x_{\rm HI} Z\, {\rm d}l$.

\begin{figure}
\centering
\includegraphics[width=9.2cm]{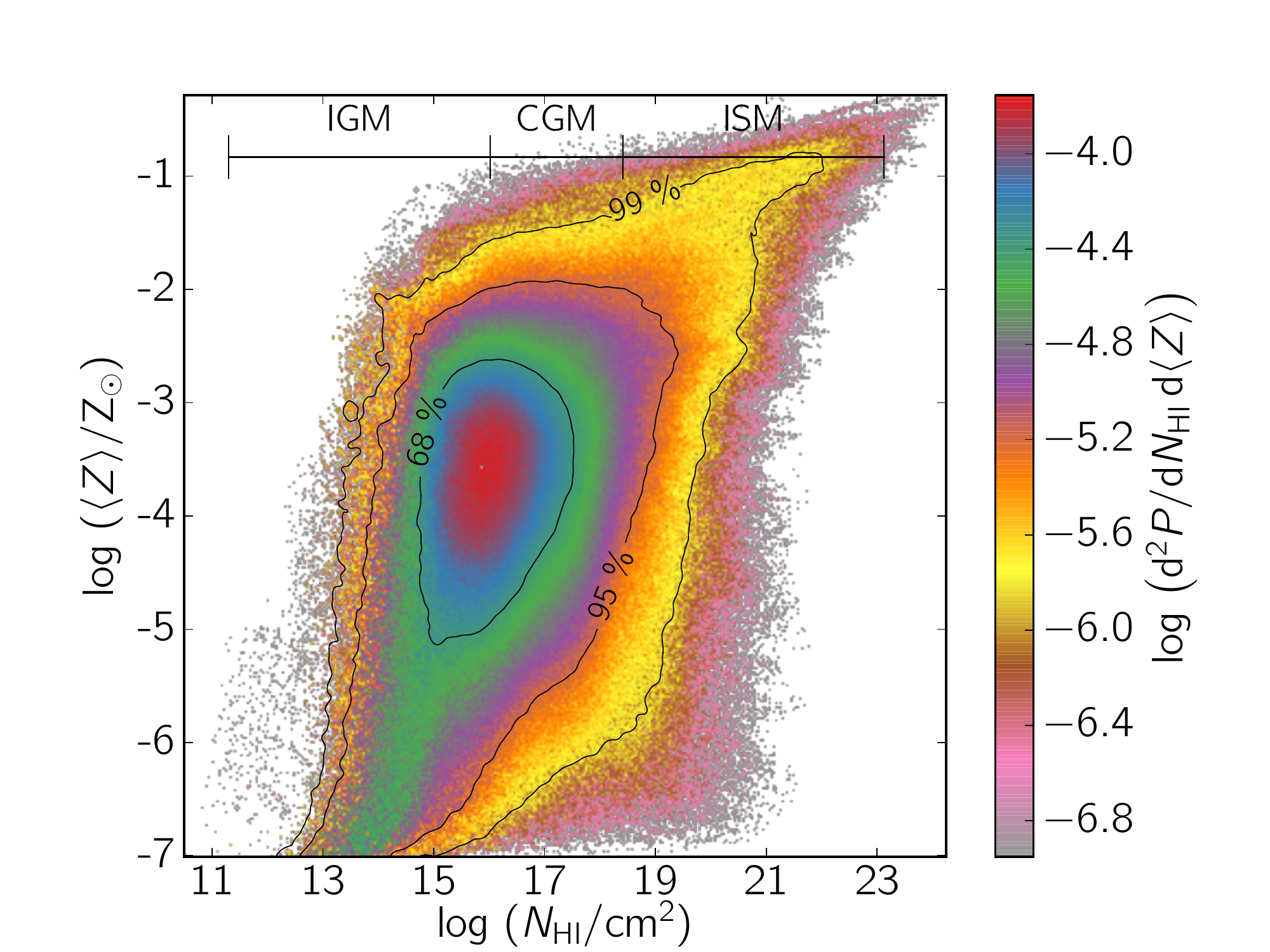}
\caption{Probability distribution function (PDF) of mean metallicity ($\zlos$) and column density ($N_{\rm HI}$) for a $M_{\halo}\simeq10^{11}\msun$ galactic environment. The color bar quantifies the PDF weighted by the l.o.s. number. The black solid lines indicate $68\%, 95\%$ and $99\%$ confidence levels.
\label{metal_column_map}}
\end{figure}

In Fig. \ref{metal_column_map}, we plot the probability distribution function (PDF) of $N_{\rm HI}$ and $\zlos$ for a simulated environment characterized by $M_{\halo}\simeq10^{11}\msun$. Consistently with observations of the CGM in the proximity of $M_{\rm h}\simeq10^{11}\msun$ halos \citep[i.e.][]{Liang:2014arXiv}, we find an upper limit of $\zlos< 10^{-1}\zsun$ in the simulated CGM.

Fig. \ref{metal_column_map} shows that $\zlos$ tightly correlates with $N_{\rm HI}$ only in the ISM, which displays both high column density ($N_{\rm HI}\gsim10^{19}{\rm cm}^{-2}$) and high mean metallicity ($\zlos\gsim10^{-1.5}\zsun$) values. However, for the CGM/IGM, the underlying tight $\Delta$-$Z$ correlation is somewhat blurred, once projected into the $N_{\rm HI}$ and $\zlos$ variables, which present larger dispersions. This result implies that \HI absorption studies do not allow to precisely constrain the CGM metallicity, as a consequence of a strong $N_{\rm HI}$ -- $\zlos$ degeneracy.

Such a degeneracy can only be broken by adding metal absorption line information. In this case, proximity effect of ionizing sources may turn out to be crucial in determining the different ionization levels of metal atoms. We defer to a future work a proper inclusion of these radiative transfer effects in our simulation. This will allow us to correctly interpret high-$z$ CGM/IGM metal absorption line observations \citep[e.g.][]{DOdorico:2013MNRAS,Gonzalo:2014arXiv}.

\section{Conclusions}\label{sec_conclusioni}
We have used a cosmological metal enrichment simulation \citep{Pallottini:2014_sim} to study the CGM/IGM properties of high-$z$ galaxies, by analyzing the \HI absorption profiles of the simulated galactic environments. The main results can be summarized as follows:
\begin{itemize}
\item[\bf 1.] At $z=4$, the gas radial density profiles in galactic environments are self-similar once scaled with the virial radius of the system, and can be fitted by a piecewise power-law ($\rho_{\rm pp}$, eq. \ref{eq_densita}). We have used this result to predict the \HI equivalent width (${\rm EW}_{\rm HI}$) as a function of the impact parameter $b$ (eq.s \ref{eq_nh} and \ref{eq_EW}).

\item[\bf 2.] Using simulations, we have produced mock \HI absorption spectra which are then analyzed using the largest gap statistics to identify CGM absorption features. As a consistency check of the ${\rm EW}_{\rm HI}$ model, we have verified that it can reproduce the analogous profile deduced from the synthetic spectra.

\item[\bf 3.] Our analytical model (calibrated at $z=4$) successfully reproduces CGM/IGM observations both at $z\simeq0$ \citep{Liang:2014arXiv} and at $z\simeq2.2$ \citep[][]{Steidel:2010ApJ}, possibly suggesting that the density profiles evolve very weakly with redshift.
\item[\bf 4.] We have investigated the relation between the mean metallicity along a simulated l.o.s. $\zlos$ and the $N_{\rm HI}$ distribution of galactic environments. Consistently with metal absorption line observations of \citet[][]{Liang:2014arXiv}, we find $\zlos< 10^{-1}\zsun$ in the CGM; however, the strong $N_{\rm HI}$ -- $\zlos$ degeneracy does not allow to constrain the CGM metallicity through \HI absorption studies alone, and metal absorption line information is required to this goal.
\end{itemize}

\section*{Acknowledgments}
We are grateful to E. Komatsu and L. Vallini for fruitful discussions.
\bibliographystyle{mn2e}
\bibliography{mass_metal_igm}
\bsp

\label{lastpage}
\end{document}

%% file: pacchetti.tex
\usepackage[english]{babel}
\usepackage{amsmath,amssymb}
\usepackage{graphicx, subfig}
\usepackage[normalem]{ulem}

\usepackage{color}
\usepackage{mathtools}
\usepackage{epstopdf}

%% file: definizioni.tex
\def\be{\begin{equation}} 
\def\ee{\end{equation}} 
\def\ba{\begin{eqnarray}} 
\def\ea{\end{eqnarray}}

\def\msun{{\Msun}}

\def\HI{\hbox{H~$\scriptstyle\rm I\ $}}

\def\gsim{\lower.5ex\hbox{\gtsima}} 
\def\lsim{\lower.5ex\hbox{\ltsima}} \def\gtsima{$\; \buildrel > \over 
\sim \;$} \def\ltsima{$\; \buildrel < \over \sim \;$} \def\prosima{$\; 
\buildrel \propto \over \sim \;$} \def\gsim{\lower.5ex\hbox{\gtsima}} 
\def\lsim{\lower.5ex\hbox{\ltsima}} 
\def\simgt{\lower.5ex\hbox{\gtsima}} 
\def\simlt{\lower.5ex\hbox{\ltsima}} 
\def\simpr{\lower.5ex\hbox{\prosima}} \def\la{\lsim} \def\ga{\gsim} 
  
 \def\gtsima{$\; \buildrel > \over \sim \;$} 
\def\ltsima{$\; \buildrel < \over \sim \;$} 
\def\gsim{\lower.5ex\hbox{\gtsima}} 
\def\lsim{\lower.5ex\hbox{\ltsima}} 
\def\simgt{\lower.5ex\hbox{\gtsima}} 
\def\simlt{\lower.5ex\hbox{\ltsima}} 
\def\simpr{\lower.5ex\hbox{\prosima}} 
\def\la{\lsim} 
\def\ga{\gsim}

\def\msun{\,{\rm \Msun}}

\def\E3{{\cal E}_{\rm g}^{III}}

\def\rvir{r_{vir}}
\def\rvir{r_{vir}}

\def\r12{r_{1/2}} 
\def\x12{x_{1/2}} 
\def\v12{v_{1/2}}

%% file: mass_metal_igm.bbl
\begin{thebibliography}{35}
\expandafter\ifx\csname natexlab\endcsname\relax\def\natexlab#1{#1}\fi

\bibitem[{{Barai} {et~al}\mbox{.}(2013){Barai}, {Viel}, {Borgani}, {Tescari},
  {Tornatore}, {Dolag}, {Killedar}, {Monaco}, {D'Odorico}, \&
  {Cristiani}}]{Barai:2013MNRAS}
{Barai} P. {et~al.}, 2013, \mnras, 430, 3213

\bibitem[{{Borthakur} {et~al}\mbox{.}(2013){Borthakur}, {Heckman},
  {Strickland}, {Wild}, \& {Schiminovich}}]{Borthakur:2013ApJ}
{Borthakur} S., {Heckman} T., {Strickland} D., {Wild} V., {Schiminovich} D.,
  2013, \apj, 768, 18

\bibitem[{{Bouwens} {et~al}\mbox{.}(2012){Bouwens}, {Illingworth}, {Oesch},
  {Franx}, {Labb{\'e}}, {Trenti}, {van Dokkum}, {Carollo}, {Gonz{\'a}lez},
  {Smit}, \& {Magee}}]{Bouwens:2012ApJ}
{Bouwens} R.~J. {et~al.}, 2012, \apj, 754, 83

\bibitem[{{Chen}(2012)}]{Chen:2012MNRAS}
{Chen} H.-W., 2012, \mnras, 427, 1238

\bibitem[{{Chen} {et~al}\mbox{.}(1998){Chen}, {Lanzetta}, {Webb}, \&
  {Barcons}}]{Chen:1998ApJ}
{Chen} H.-W., {Lanzetta} K.~M., {Webb} J.~K., {Barcons} X., 1998, \apj, 498, 77

\bibitem[{{Chen} {et~al}\mbox{.}(2001){Chen}, {Lanzetta}, {Webb}, \&
  {Barcons}}]{Chen:2001ApJ}
{Chen} H.-W., {Lanzetta} K.~M., {Webb} J.~K., {Barcons} X., 2001, \apj, 559,
  654

\bibitem[{{Churchill} {et~al}\mbox{.}(2013){Churchill}, {Trujillo-Gomez},
  {Nielsen}, \& {Kacprzak}}]{Churchill:2013arXiv}
{Churchill} C.~W., {Trujillo-Gomez} S., {Nielsen} N.~M., {Kacprzak} G.~G.,
  2013, \apj, 779, 87

\bibitem[{{Crain} {et~al}\mbox{.}(2013){Crain}, {McCarthy}, {Schaye}, {Theuns},
  \& {Frenk}}]{Crain:2013MNRAS}
{Crain} R.~A., {McCarthy} I.~G., {Schaye} J., {Theuns} T., {Frenk} C.~S., 2013,
  \mnras, 432, 3005

\bibitem[{{Croft}(1998)}]{Croft:1998tsra.conf}
{Croft} R.~A.~C., 1998, in Eighteenth Texas Symposium on Relativistic
  Astrophysics, {Olinto} A.~V., {Frieman} J.~A., {Schramm} D.~N., eds., p. 664

\bibitem[{{Dayal} {et~al}\mbox{.}(2008){Dayal}, {Ferrara}, \&
  {Gallerani}}]{Dayal:2008MNRAS}
{Dayal} P., {Ferrara} A., {Gallerani} S., 2008, \mnras, 389, 1683

\bibitem[{{D'Odorico} {et~al}\mbox{.}(2013){D'Odorico}, {Cupani}, {Cristiani},
  {Maiolino}, {Molaro}, {Nonino}, {Centuri{\'o}n}, {Cimatti}, {di Serego
  Alighieri}, {Fiore}, {Fontana}, {Gallerani}, {Giallongo}, {Mannucci},
  {Marconi}, {Pentericci}, {Viel}, \& {Vladilo}}]{DOdorico:2013MNRAS}
{D'Odorico} V. {et~al.}, 2013, \mnras, 435, 1198

\bibitem[{{Finlator} {et~al}\mbox{.}(2013){Finlator}, {Mu{\~n}oz},
  {Oppenheimer}, {Oh}, {{\"O}zel}, \& {Dav{\'e}}}]{Finlator:2013MNRAS}
{Finlator} K., {Mu{\~n}oz} J.~A., {Oppenheimer} B.~D., {Oh} S.~P., {{\"O}zel}
  F., {Dav{\'e}} R., 2013, \mnras, 436, 1818

\bibitem[{{Gallerani} {et~al}\mbox{.}(2006){Gallerani}, {Choudhury}, \&
  {Ferrara}}]{Gallerani:2006MNRAS}
{Gallerani} S., {Choudhury} T.~R., {Ferrara} A., 2006, \mnras, 370, 1401

\bibitem[{{Gallerani} {et~al}\mbox{.}(2008{\natexlab{a}}){Gallerani},
  {Ferrara}, {Fan}, \& {Choudhury}}]{Gallerani:2008MNRASa}
{Gallerani} S., {Ferrara} A., {Fan} X., {Choudhury} T.~R., 2008{\natexlab{a}},
  \mnras, 386, 359

\bibitem[{{Gallerani} {et~al}\mbox{.}(2008{\natexlab{b}}){Gallerani},
  {Salvaterra}, {Ferrara}, \& {Choudhury}}]{Gallerani:2008MNRASb}
{Gallerani} S., {Salvaterra} R., {Ferrara} A., {Choudhury} T.~R.,
  2008{\natexlab{b}}, \mnras, 388, L84

\bibitem[{{Gonz{\'a}lez} {et~al}\mbox{.}(2011){Gonz{\'a}lez}, {Labb{\'e}},
  {Bouwens}, {Illingworth}, {Franx}, \& {Kriek}}]{Gonzalez:2011}
{Gonz{\'a}lez} V., {Labb{\'e}} I., {Bouwens} R.~J., {Illingworth} G., {Franx}
  M., {Kriek} M., 2011, \apjl, 735, L34

\bibitem[{{Gonzalo D{\'{\i}}az} {et~al}\mbox{.}(2014){Gonzalo D{\'{\i}}az},
  {Koyama}, {Ryan-Weber}, {Cooke}, {Ouchi}, {Shimasaku}, \&
  {Nakata}}]{Gonzalo:2014arXiv}
{Gonzalo D{\'{\i}}az} C., {Koyama} Y., {Ryan-Weber} E.~V., {Cooke} J., {Ouchi}
  M., {Shimasaku} K., {Nakata} F., 2014, ArXiv:1404.7656

\bibitem[{{Haardt} \& {Madau}(2012)}]{Haardt:2012}
{Haardt} F., {Madau} P., 2012, \apj, 746, 125

\bibitem[{{Iapichino} {et~al}\mbox{.}(2013){Iapichino}, {Viel}, \&
  {Borgani}}]{Iapichino:2013MNRAS}
{Iapichino} L., {Viel} M., {Borgani} S., 2013, \mnras, 432, 2529

\bibitem[{{Jia Liang} \& {Chen}(2014)}]{Liang:2014arXiv}
{Jia Liang} C., {Chen} H.-W., 2014, ArXiv 1402.3602

\bibitem[{{Keating} {et~al}\mbox{.}(2014){Keating}, {Haehnelt}, {Becker}, \&
  {Bolton}}]{Keating:2013arXiv}
{Keating} L.~C., {Haehnelt} M.~G., {Becker} G.~D., {Bolton} J.~S., 2014,
  \mnras, 438, 1820

\bibitem[{{Khandai} {et~al}\mbox{.}(2014){Khandai}, {Di Matteo}, {Croft},
  {Wilkins}, {Feng}, {Tucker}, {DeGraf}, \& {Liu}}]{Khandai:2014arXiv}
{Khandai} N., {Di Matteo} T., {Croft} R., {Wilkins} S.~M., {Feng} Y., {Tucker}
  E., {DeGraf} C., {Liu} M.-S., 2014, ArXiv 1402.0888

\bibitem[{{Larson} {et~al}\mbox{.}(2011){Larson}, {Dunkley}, {Hinshaw},
  {Komatsu}, {Nolta}, {Bennett}, {Gold}, {Halpern}, {Hill}, {Jarosik}, {Kogut},
  {Limon}, {Meyer}, {Odegard}, {Page}, {Smith}, {Spergel}, {Tucker}, {Weiland},
  {Wollack}, \& {Wright}}]{Larson:2011}
{Larson} D. {et~al.}, 2011, \apjs, 192, 16

\bibitem[{{Meiksin}(2009)}]{Meiksin:2009RvMP}
{Meiksin} A.~A., 2009, Reviews of Modern Physics, 81, 1405

\bibitem[{{Nielsen} {et~al}\mbox{.}(2013){Nielsen}, {Churchill}, \&
  {Kacprzak}}]{Nielsen:2013ApJ}
{Nielsen} N.~M., {Churchill} C.~W., {Kacprzak} G.~G., 2013, \apj, 776, 115

\bibitem[{{Oppenheimer} {et~al}\mbox{.}(2012){Oppenheimer}, {Dav{\'e}}, {Katz},
  {Kollmeier}, \& {Weinberg}}]{Oppenheimer:2012MNRAS}
{Oppenheimer} B.~D., {Dav{\'e}} R., {Katz} N., {Kollmeier} J.~A., {Weinberg}
  D.~H., 2012, \mnras, 420, 829

\bibitem[{{Pallottini} {et~al}\mbox{.}(2014){Pallottini}, {Ferrara},
  {Gallerani}, {Salvadori}, \& {D'Odorico}}]{Pallottini:2014_sim}
{Pallottini} A., {Ferrara} A., {Gallerani} S., {Salvadori} S., {D'Odorico} V.,
  2014, \mnras, 440, 2498

\bibitem[{{Pieri} {et~al}\mbox{.}(2013){Pieri}, {Mortonson}, {Frank},
  {Crighton}, {Weinberg}, {Lee}, {Noterdaeme}, {Bailey}, {Busca}, {Ge},
  {Kirkby}, {Lundgren}, {Mathur}, {Paris}, {Palanque-Delabrouille},
  {Petitjean}, {Rich}, {Ross}, {Schneider}, \& {York}}]{Pieri:2013arXiv}
{Pieri} M.~M. {et~al.}, 2013, ArXiv 1309.6768

\bibitem[{{Rudie} {et~al}\mbox{.}(2013){Rudie}, {Steidel}, {Shapley}, \&
  {Pettini}}]{Rudie:2013ApJ}
{Rudie} G.~C., {Steidel} C.~C., {Shapley} A.~E., {Pettini} M., 2013, \apj, 769,
  146

\bibitem[{{Rudie} {et~al}\mbox{.}(2012){Rudie}, {Steidel}, {Trainor}, {Rakic},
  {Bogosavljevi{\'c}}, {Pettini}, {Reddy}, {Shapley}, {Erb}, \&
  {Law}}]{Rudie:2012ApJ}
{Rudie} G.~C. {et~al.}, 2012, \apj, 750, 67

\bibitem[{{Shen} {et~al}\mbox{.}(2013){Shen}, {Madau}, {Guedes}, {Mayer},
  {Prochaska}, \& {Wadsley}}]{Shen:2013ApJ}
{Shen} S., {Madau} P., {Guedes} J., {Mayer} L., {Prochaska} J.~X., {Wadsley}
  J., 2013, \apj, 765, 89

\bibitem[{{Shull}(2014)}]{Shull:2014arXiv}
{Shull} J.~M., 2014, \apj, 784, 142

\bibitem[{{Songaila}(2004)}]{Songaila:2004AJ}
{Songaila} A., 2004, \aj, 127, 2598

\bibitem[{{Steidel} {et~al}\mbox{.}(2010){Steidel}, {Erb}, {Shapley},
  {Pettini}, {Reddy}, {Bogosavljevi{\'c}}, {Rudie}, \&
  {Rakic}}]{Steidel:2010ApJ}
{Steidel} C.~C., {Erb} D.~K., {Shapley} A.~E., {Pettini} M., {Reddy} N.,
  {Bogosavljevi{\'c}} M., {Rudie} G.~C., {Rakic} O., 2010, \apj, 717, 289

\bibitem[{{Teyssier}(2002)}]{Teyssier:2002}
{Teyssier} R., 2002, \aap, 385, 337

\end{thebibliography}
